\documentclass[sigconf,nonacm,natbib=true]{acmart}

\usepackage{booktabs}
\usepackage{pbalance}
\usepackage{listings}
\usepackage{xcolor}
\usepackage{subcaption}
\usepackage{graphicx}

\setcopyright{none}
\renewcommand\footnotetextcopyrightpermission[1]{}
\acmConference[CIKM '26]{Proceedings of the 35th ACM International Conference on Information and Knowledge Management}{November 9--11, 2026}{Rome, Italy}
\acmYear{2026}
\copyrightyear{2026}

\title[OpenIIR: An Open Simulation Platform for Information Retrieval Research]{\textsc{OpenIIR}: An Open Simulation Platform \\ for Information Retrieval Research}

\author{Saber Zerhoudi}
\affiliation{%
  \institution{University of Passau}
  \city{Passau}
  \country{Germany}
}
\email{szerhoudi@acm.org}

\makeatletter
\newcommand{\acmrightssize}{\fontsize{8}{9.5}\selectfont}

\newcommand{\firstpagerights}[1]{%
  \begingroup
    \renewcommand\thefootnote{}%
    \footnotetext{%
      \acmrightssize
      \raggedright
      \setlength{\parskip}{0pt}%
      \setlength{\parindent}{0pt}%
      #1%
    }%
    \addtocounter{footnote}{0}%
  \endgroup
}
\makeatother

\begin{abstract}
\textbf{OpenIIR}\footnote{\href{https://openiir.com}{https://openiir.com}} runs hundreds of LLM-driven personas as parameterised, reproducible IR research experiments. Researchers configure agents across four kinds of multi-agent study (deliberative panels, social platforms, curated recommender feeds, and evolutionary co-evolution between content producers and credibility detectors) under many priors, rounds, and constraints. Persona budgets, retrieval policies, ranker choices, intervention timings, and mutation rates are declared up front, and the same study can be re-run under different settings to compare outcomes side by side. Every run produces structured outputs (argument graphs, exposure logs, fitness traces, transcripts) that a downstream evaluator can consume directly, and a new study is a 200--400 line plug-in over a shared core (agent runtime, world-model store, retrieval primitives, claim extractor, persona ontology). The contributions are: (i)~the shared core; (ii)~a type interface for pluggable scenarios; (iii)~four released types with reference runs (\emph{Panel}, \emph{Social-Media}, \emph{Curated-Feed}, \emph{Multi-Generational}); and (iv)~six modular extensions sketched against open IR research questions. 
\end{abstract}


\begin{document}


\maketitle
\enlargethispage{2\baselineskip}
\firstpagerights{%
  © ACM, 2026. This is the author's version of the work.\\
}

\section{Introduction}
\label{sec:intro}

\textbf{OpenIIR} lets IR researchers run controlled experiments on agent-driven information phenomena. Hundreds of LLM-driven persona agents can be configured to deliberate in budgeted panels, post and propagate on a social platform, scroll through a curated feed, or co-evolve as content producers and credibility detectors across many generations. Every run is parameterised (persona priors, retrieval budgets, ranker choices, intervention timings, mutation rates, stopping conditions), reproducible end-to-end, and produces structured outputs (argument graphs, exposure logs, fitness traces, transcripts) that a downstream evaluator can consume without bespoke parsing.

The kinds of question this enables are difficult to answer otherwise. How would a workshop's panellists actually disagree on a given theme, when no schedule can put them all in one room? How does a recommender shift opinion variance across a user population over twelve simulated weeks, when the only available alternative is an A/B test that ships to real users? How fast can content producers stay ahead of credibility detectors over fifty generations, when running fifty real generations would take years? OpenIIR collapses each of these into experiments that run on a single laptop in minutes to hours, and that can be re-run under different priors, budgets, and interventions to see what changes~\cite{zerhoudi2026guided,balog2025theory,behind-prompt}.

The platform is built around a small shared core (agent runtime, world-model store, retrieval primitives, claim extractor, persona ontology) and a typed interface that pluggable scenario types implement. A type defines its own platform actions, scheduling, metrics, and audience-facing surfaces; the core is shared, so a new study becomes a 200--400 line plug-in. Four types are released. \emph{Panel} runs budgeted persona agents through structured deliberative rounds and produces an argument graph as the structured output; this type was used at ECIR~2026 Search Futures~\cite{azzopardi2026third} to run four parallel breakout panels with workshop-participant personas. \emph{Social-Media} runs posting, reposting, liking, and commenting on a Twitter- or Reddit-like platform with persona-conditioned activity profiles. \emph{Curated-Feed} drives a recommender against agent users with belief vectors. \emph{Multi-Generational} runs producer-detector co-evolution. Existing simulation platforms each cover one slice of this space: OASIS~\cite{oasis}, Concordia~\cite{concordia}, Generative Agents~\cite{generative-agents}, SimIIR~2.0 and 3~\cite{simiir2,simiir3}. None of them share an agent runtime, a world-model store, and an evaluation surface so that the same research team can move between settings without rebuilding the foundations every time. Six modular extensions to the platform are sketched in Table~\ref{tab:roadmap}, each a possible scenario type that another research team could plug in over the same shared core.

The contributions of this paper are: (i)~a reusable shared core for IR simulation; (ii)~a type interface that makes a new scenario a small plug-in; (iii)~four released types with worked examples and reference runs; (iv)~a use case of the Panel type at ECIR~2026 Search Futures, included as one example of the platform in use; and (v)~the modular extension space sketched in Table~\ref{tab:roadmap}, with each row tied to a concrete IR research question.

\section{Related Work}
\label{sec:related}

Three families of prior work are closest to OpenIIR.

\textit{Multi-agent simulation platforms.} OASIS~\cite{oasis} simulates social-platform interactions at the scale of one million Twitter or Reddit personas with persona-conditioned action policies. Concordia~\cite{concordia} provides a deliberation simulator with grounded action spaces. Generative Agents~\cite{generative-agents} simulates a small town with episodic memory and reflection. SOTOPIA~\cite{sotopia} benchmarks social interaction tasks for language agents. Domain instantiations of generative agents have been deployed in IR-adjacent settings, for example digital libraries~\cite{generative-agents-dl}. Each of these platforms is purpose-built for a single setting. OpenIIR contributes a shared core (agent runtime, world-model store, claim extractor, persona ontology) plus a typed interface that lets the same primitives drive a deliberation, a feed-curation study, and a generational arms race without rebuilding infrastructure.

\textit{Synthetic users for IR evaluation.} A long line of work simulates IR users to support evaluation at scale. SimIIR 2.0~\cite{simiir2} and SimIIR 3~\cite{simiir3} provide Markov-model and conversational user simulators for IIR. Recent generative-AI variants include USimAgent~\cite{usimagent} for search-session simulation and the Faggioli et al. study of LLM relevance judgement~\cite{faggioli2023perspectives}. The SIGIR 2025 tutorial of Balog et al.~\cite{balog2025theory} surveys the design space, and the Sim4IA workshop series~\cite{sim4ia2025} maintains the community context. OpenIIR's Curated-Feed type generalises this from per-query relevance to longitudinal exposure, which is the regime where invisible curation effects accumulate.

\textit{Information ecosystem modelling.} Compartmental models such as the RGB framework~\cite{rgb} characterise high-level information dynamics with rate parameters and bias terms. Model-collapse work~\cite{model-collapse} studies the consequences of training generative models on their own outputs. Empirical studies of diffusion such as Vosoughi et al.~\cite{vosoughi2018spread} and exposure on social platforms such as Bakshy et al.~\cite{bakshy2015exposure} ground the questions OpenIIR's typed scenarios target. OpenIIR cannot retrain its underlying LLM, so the Multi-Generational type evolves prompt templates and retrieval policies across generations, which operationalises the same questions at the agent layer.

\section{Platform Architecture}
\label{sec:arch}

\begin{figure}[t]
    \centering
    \includegraphics[width=1\linewidth]{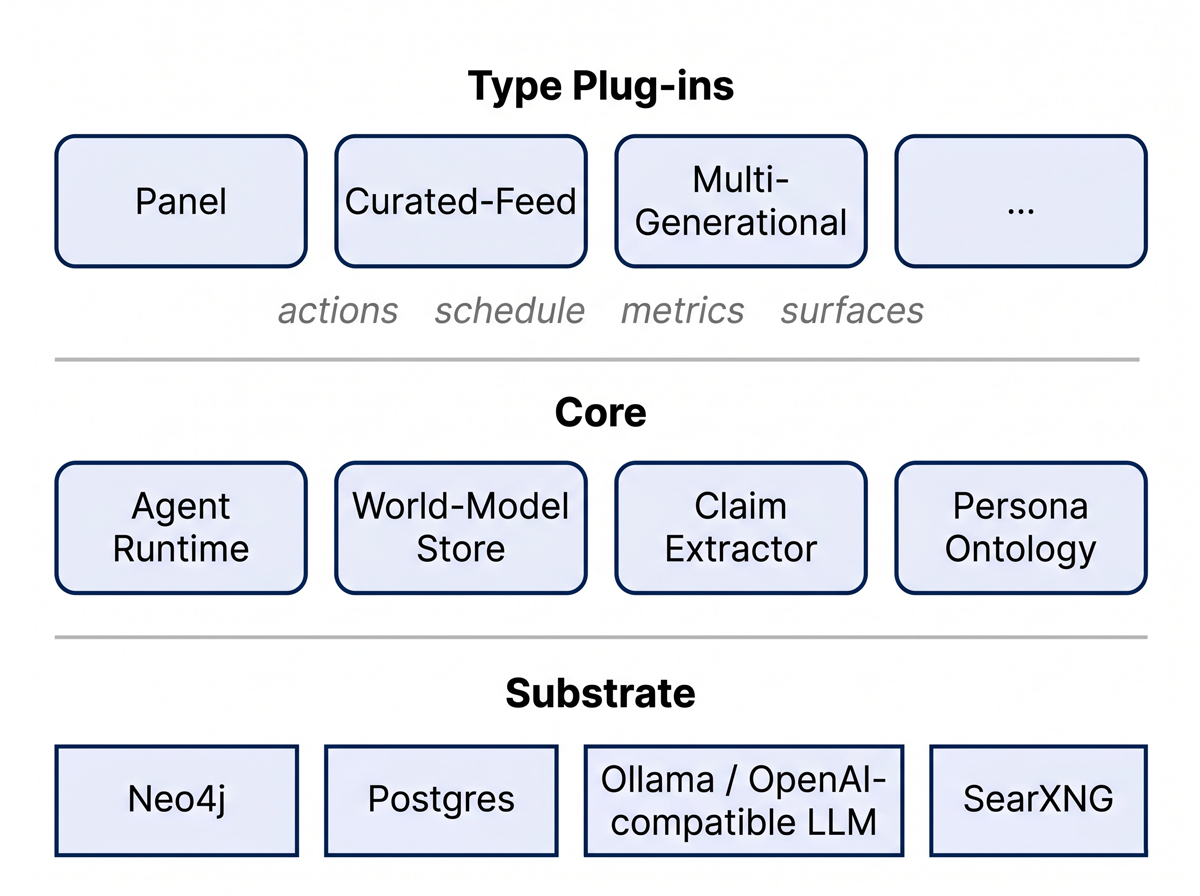}
    \caption{Three-layer architecture. The core is type-agnostic. The substrate is local-first by default; cloud LLMs are a one-line config swap.}
    \label{fig:arch}
\end{figure}

OpenIIR is built around a small core. Anything that several types share lives there. Anything that is specific to one scenario lives in the type plug-in. Figure~\ref{fig:arch} shows the layering.

\paragraph{Agent runtime.}
A persona is a one-paragraph bio plus structured fields (token budget, search budget, optional stance and influence weight, optional activity profile). Personas can be authored by hand or extracted from a world model by a one-pass generator that the core provides. Budgets are debited as the agent acts and are exposed to type plug-ins so they can decide when to skip an agent or end a round.

\paragraph{World-model store.}
Each scenario carries its own world model: a chunked corpus, a keyword-lookup tool over the chunks, and an optional justified-search tool that calls SearXNG or a configured external API. A search query without an attached justification is rejected before any budget is debited. World models are isolated per scenario instance, so two simulations sharing personas do not share corpus state.

\paragraph{Retrieval primitives.}
The core ships hybrid lexical-vector search (BM25 plus embeddings) over the world model with a configurable mixing weight. Type plug-ins use this for any retrieval their scenario requires, including agent lookups, recommender candidate generation, and post-hoc interrogation.

\paragraph{Claim extractor.}
After each agent utterance or post, a separate extractor reads the new text, tags any new claim with its stance towards the topic (supporting, challenging, neutral) and links it back to earlier claims with typed edges (\emph{supports}, \emph{counters}, \emph{refines}, \emph{questions}). The graph of claims is stored alongside the transcript and is the primary structured output a downstream evaluator consumes.

\paragraph{Persona ontology.}
A persona library survives across types. The same researcher persona used in a Panel can be re-used as one of the agent users in a Curated-Feed run, or as a producer or detector in a Multi-Generational run. The ontology fields tracked are scenario-agnostic.

\paragraph{Type interface.}
A new type implements four methods: \texttt{actions()} returns the platform actions agents may take; \texttt{schedule()} drives turns, rounds, or generations; \texttt{metrics()} reports type-specific evaluation; \texttt{surfaces()} returns the live views and post-hoc interrogation endpoints. A reference type plug-in is around 250 lines.

\section{Released Types}
\label{sec:types}

\subsection{Panel}
\label{sec:panel}

The Panel type runs a small group of personas through structured deliberative rounds. A round has a name, a goal sentence, and a hard cap on the number of turns. An LLM-driven moderator decides at each turn who speaks next and what they should address. The same moderator may advance the discussion to the next round or end the panel once the round's goal is met. Personas may call the keyword-lookup tool freely; web search requires a written justification, and an under-justified query is rejected before any search budget is debited. Once a persona's token budget is exhausted, the moderator skips that persona on subsequent turns, which prevents a single panelist from dominating the discussion.

\begin{figure*}[t]
    \centering
    \includegraphics[width=\linewidth]{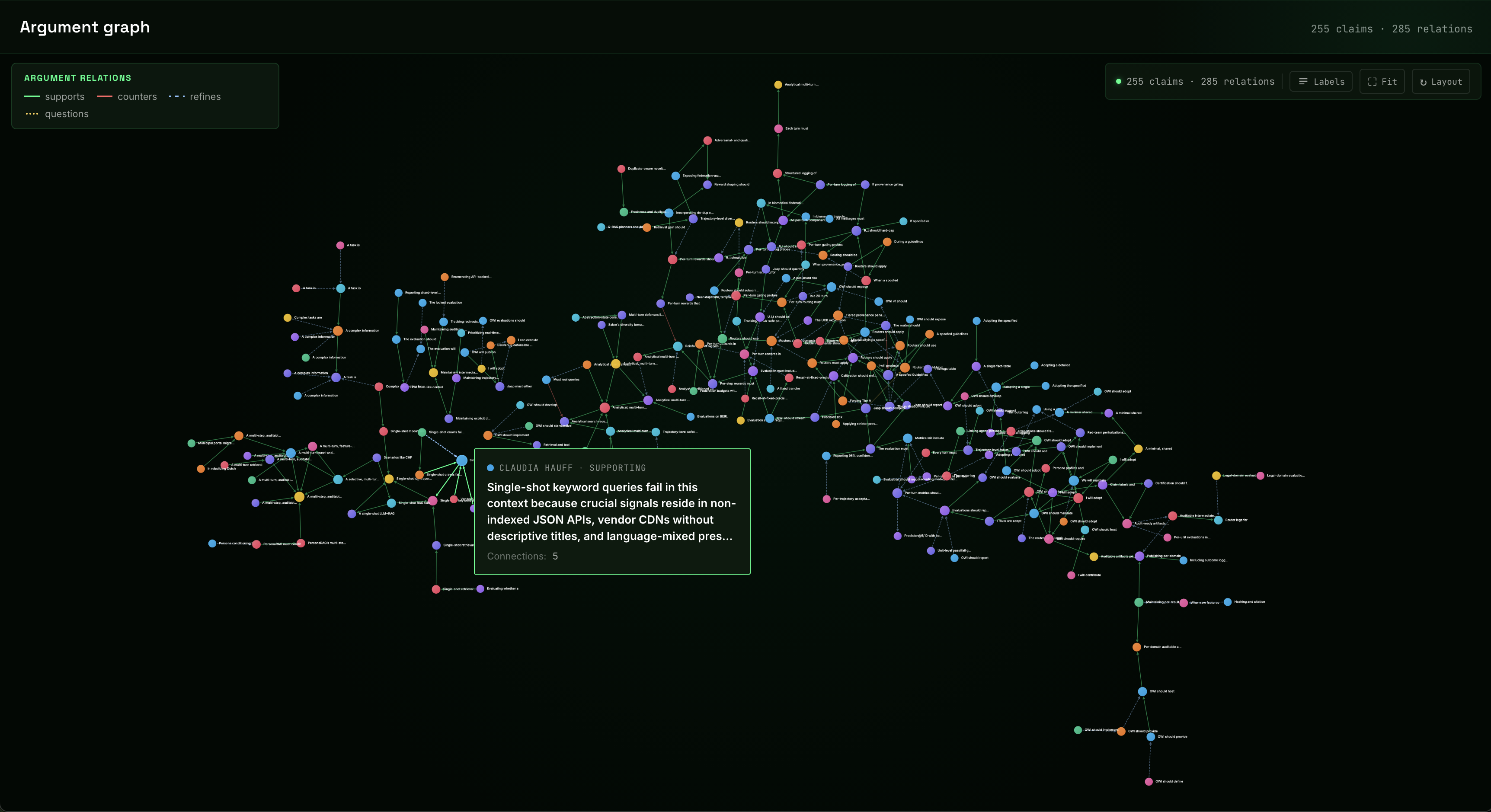}
    \caption{Panel argument graph: claim nodes connected by typed edges (\emph{supports}, \emph{counters}, \emph{refines}, \emph{questions}). Screenshot from a live run at \texttt{openiir.com/simulation/sim\_0d9f423da0ac}.}
    \label{fig:panel-arg}
\end{figure*}

\begin{figure}[t]
    \centering
    \includegraphics[width=\linewidth]{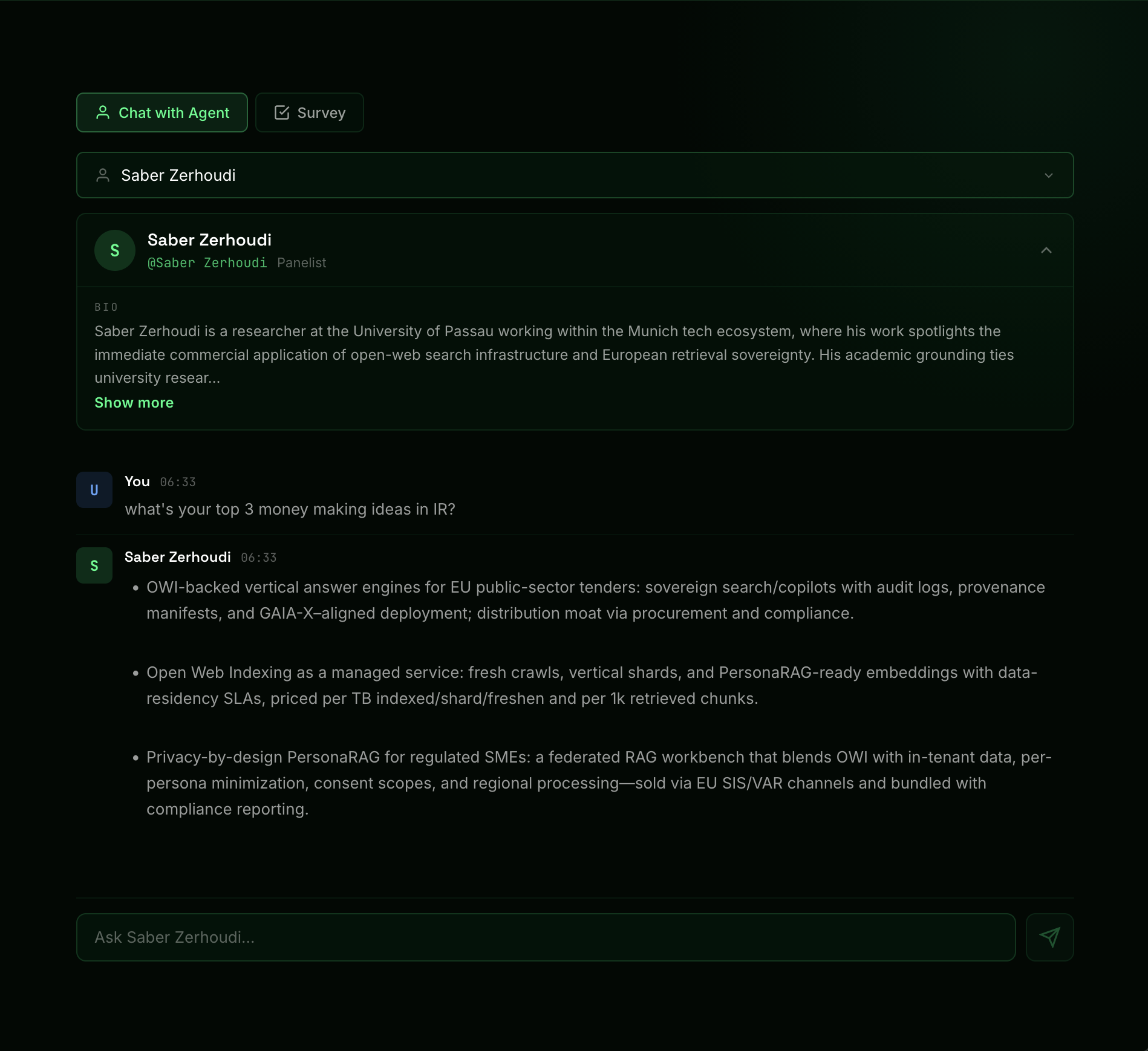}
    \caption{Per-panelist chat: converse with a single panelist whose context is restricted to their own utterances and claims, or pose one question to every panelist at once.}
    \label{fig:panel-chat}
\end{figure}

\begin{figure}[t]
    \centering
    \includegraphics[width=\linewidth]{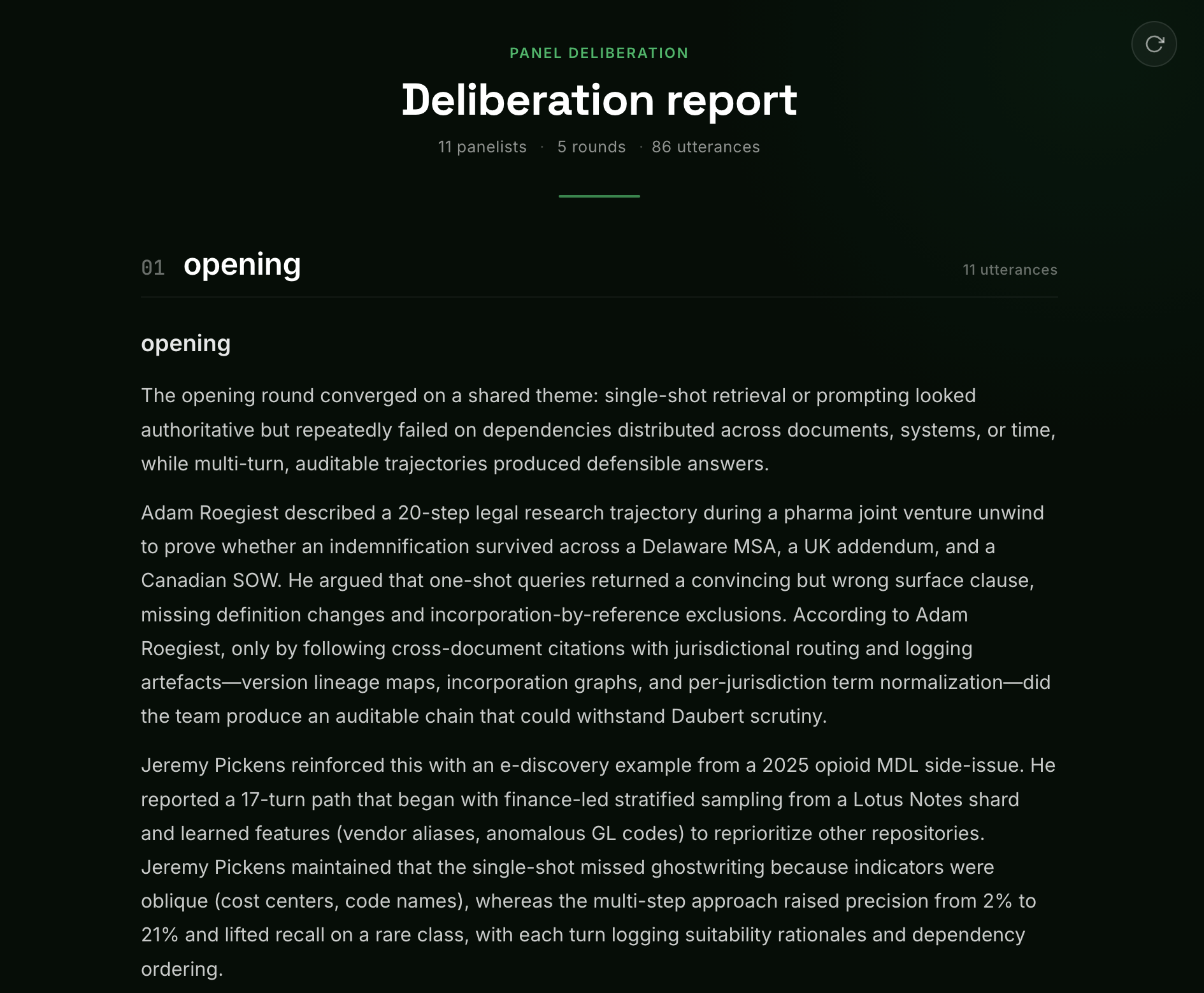}
    \caption{Panel deliberation report: synthesis of positions, convergence, and unresolved disagreements.}
    \label{fig:panel-report}
\end{figure}

The argument graph is the type's structured retrieval output. After every utterance, a separate extractor reads the new text, tags any new claim with its stance towards the topic (supporting, challenging, neutral), and links it back to earlier claims with typed edges (\emph{supports}, \emph{counters}, \emph{refines}, \emph{questions}). Figure~\ref{fig:panel-arg} shows the resulting argument graph for one of the ECIR~2026 SF3 panels. Beyond the chronological transcript, two further surfaces consume the graph: a per-panelist chat (Figure~\ref{fig:panel-chat}) that lets one converse with a single panelist whose context is restricted to their own utterances and claims, or pose one question to every panelist at once, and a deliberation report (Figure~\ref{fig:panel-report}) that summarises positions, convergence, and unresolved disagreements.

Three round shapes ship with the type: the standard \emph{opening / deliberation / wrap-up} shape used for most workshop breakouts, a \emph{Delphi} shape with revision turns suitable for synthetic expert panels, and a \emph{pitch-storm / kill-or-keep / final-ten} shape for generative ideation. A new shape is a small declarative configuration; no code change is needed.

\subsection{Social-Media}
\label{sec:social}

The Social-Media type runs persona-driven posting and propagation on a Twitter- or Reddit-like platform. Agents may post, repost, comment, like, dislike, follow, and search past posts. Each persona carries an activity profile (posts per hour, comments per hour, active hours of day, response delay), an influence weight that scales the visibility of their posts to other agents, and an optional sentiment bias and stance. The schedule advances in discrete rounds (default sixty simulated minutes per round), with optional milestone events that can shift activity rates or inject content at specified rounds.

The platform layer is a thin abstraction so that the same simulation can target either Twitter-style or Reddit-style profile generators, the difference being the per-persona attributes the generator emits. The world model seeds initial topics and source content, and agent posts are written back to it through the claim extractor, so propagation can be tracked through the typed-edge graph and via the post log. Use cases supported with the released configuration include misinformation cascade studies, persona-conditioned engagement audits, and counter-speech intervention experiments.

\subsection{Curated-Feed}
\label{sec:curated}

The Curated-Feed type targets a question that recurs across the IR research agenda: can helpful but invisible curation flatten opinion variance even when no adversarial intent is present~\cite{bakshy2015exposure}? The platform is a feed of items drawn from the world model. Each agent user $u$ carries a belief vector $b_u \in \mathbb{R}^k$ over a small topic taxonomy of size $k$ (typically 8 to 16 topics), a click model $p(\text{click} \mid b_u, \text{item})$ conditioned on belief alignment, and an update rule $b_u \leftarrow b_u + \eta \cdot \Delta(\text{exposed}, \text{clicked})$ that nudges the belief vector after each exposure with learning rate $\eta$. The cognitive-state design follows our earlier work on cognitive-aware user simulation~\cite{cognitive-aware}.

A configurable ranker scores candidates per user. The type ships three rankers. \emph{Popularity} ranks by global click count and serves as the non-personalised baseline. \emph{Similarity-to-belief} ranks candidates by cosine similarity to $b_u$ and serves as the personalisation baseline. \emph{Similarity-to-helpful-history} reranks candidates by alignment to the user's most-engaged-with past items and is a stylised proxy for an engagement-aligned recommender~\cite{personarag}. A custom ranker is a single Python callable registered through the type interface.

Metrics include opinion variance across the user population, exposure entropy per topic, Kendall's~$\tau$ between an oracle ranking (ground-truth alignment with $b_u$) and the realised ranking, and per-topic exposure share. The release ships a reference run of 200 user agents over 12 simulated weeks; the run logs are exported as Parquet with per-impression features (user, item, topic, ranker score, oracle score, click) so that offline counterfactual analyses such as inverse-propensity scoring can be run without re-executing the simulation.

\subsection{Multi-Generational}
\label{sec:multigen}

The Multi-Generational type targets the question of whether content producers can stay ahead of credibility detectors across many generations~\cite{vosoughi2018spread}. Two interacting populations are simulated. \emph{Producers} author content with a fitness function on landed-claim survival in the corpus. \emph{Detectors} score new content with a fitness function on calibrated rejection of false claims. Each generation runs an evaluation match-up on a shared eval set, computes fitness per agent, samples the next generation with mutation, and reports per-generation summaries.

Mutation operates on three loci: the prompt template, the retrieval-policy parameters used during content generation or detection, and the persona bio. The underlying LLM is not retrained, and the design intentionally restricts adaptation to artefacts a researcher can inspect after a run. The eval set is built from a paired-document audit~\cite{trec-health} where every item has plausibility-matched true and false versions of the same claim, which makes producer landing rate and detector calibration directly comparable across generations.

The release ships a 50-generation reference run with $|P|=20$ producers and $|D|=20$ detectors. Lineage is tracked, so an attendee watching the live demo can trace a winning producer template back to the generation where its mutation was introduced and observe the detector adaptations that followed. The platform stores every cohort's prompts, fitness scores, and eval outcomes in a single SQLite snapshot, so a researcher can replay a published reference run on their own laptop without issuing any LLM call.

\begin{table*}[t]
\centering
\small
\caption{Modular extensions of the platform. Each row sketches a scenario type that could be plugged in over the same shared core, motivated by a concrete IR research question; the right column names the new primitives the type would add.}
\label{tab:roadmap}
\begin{tabular}{p{0.34\linewidth}p{0.18\linewidth}p{0.40\linewidth}}
\toprule
\textbf{IR question} & \textbf{Type} & \textbf{New primitives needed} \\
\midrule
What collapses first when agents stop clicking on search results? & Feedback-Outage & Toggleable signal channels, downstream evaluators, cascade trace \\
Does regulating search produce better outcomes than leaving it alone? & Twin-Ecosystem & Two world models in lockstep, policy hooks on retrieval and ranking, paired comparator metrics \\
Do persistent-memory agents gain unfair advantage over stateless ones? & Asymmetric-Memory & Per-agent memory policy, memory-access logging, advantage metrics \\
Does agent-mediated search create a new information class divide? & Tiered-Capability & Agent tiers with different retrieval primitives, equity metrics \\
What funds the open web when agents skip ads? Do humans retreat from agent-dominated spaces? & Open-Web & Trust dynamics across agent and human authors, follower graph with influence accrual, optional economic layer \\
Do agent searchers form spontaneous information cartels? & Coordination-Emergent & Independent-goal scaffolding, source-overlap and access-hoarding metrics \\
\bottomrule
\end{tabular}
\end{table*}

\section{Use Case: ECIR 2026 Search Futures}
\label{sec:ecir}

OpenIIR's Panel type was used to run four parallel breakout panels at the ECIR~2026 Search Futures workshop~\cite{azzopardi2026third}. Prior editions of the workshop~\cite{azzopardi2024report,clarke2025report} convened breakouts physically: small groups of attendees self-organised around tables and a rapporteur produced a written summary afterwards. Search Futures 3 replaced four physical breakouts with four OpenIIR Panel runs, each grounded in its own theme and its own world model, sharing a cast of eleven LLM-driven persona agents. The persona bios were drafted from each participant's recent papers and talks, then reviewed by the participant before the panel began.

The world model for each theme was assembled with a deep-research pass and seeded with the resulting synthesised report plus every cited PDF and web article. OpenIIR chunked the world model and exposed it through the keyword-lookup tool. Three of the four panels used the standard opening / deliberation / wrap-up shape. Theme~3 used the pitch-storm / kill-or-keep / final-ten shape for a generative session on commercial futures.

After every utterance, the claim extractor tagged any new claim with its stance and linked it to earlier claims with typed edges. The argument graph was used by the workshop organisers as a sanity check that the synthesised report captured the major disagreements, not only the points of convergence. Each panel ran to roughly thirty utterances over three rounds and finished in about fifteen minutes. The transcripts, argument graphs, persona ontologies, agent rosters, synthesised reports, and per-panelist interrogation surfaces are public at \texttt{openiir.com/simulation/}.

A practical consequence of this format, which physical breakouts do not allow, is that every workshop participant could be present in every panel. Rather than picking a single table, each participant was simulated across all four themes through their own persona. The four breakout reports in the workshop proceedings document the resulting deliberations, including the disagreements, side by side.

\section{Implementation and Reproducibility}
\label{sec:impl}

OpenIIR is implemented as a Flask backend, a Vue frontend, and a small set of supporting services. The graph store is Neo4j 5.15 Community Edition. Logs and exposure data go to Postgres. The default LLM is served through Ollama using \texttt{qwen3.5:32b}\cite{yang2025qwen3} for moderation and persona reasoning and \texttt{nomic-embed-text}\cite{nussbaum2024nomic} for embeddings, but any OpenAI-compatible endpoint works as a one-line config change. Justified web search is routed through SearXNG by default, with Tavily as an opt-in alternative.

The full stack runs in Docker Compose on a single host. We document three hardware tiers. A 32~GB RAM, 24~GB VRAM workstation runs the full \texttt{qwen2.5:32b} pipeline at roughly 12~min per 30-utterance Panel and 7~min per 50-generation Multi-Generational replay. A 16~GB RAM, 8~GB VRAM laptop runs \texttt{qwen2.5:7b} with reduced fidelity, sufficient for live demos. CPU-only mode is supported but slow for an audience.

For reproducibility, the platform records, per simulation, the LLM model name and version, every system prompt, every world-model chunk hash, every random seed, every search query and its justification, and the full transcript of LLM calls. A run can be replayed deterministically against a cached call log without re-issuing requests. Exported runs are signed with a content hash and published at \texttt{openiir.com/simulation/<id>}, so a reviewer can verify that a published transcript matches the released artefact.

\section{Limitations and Extension Space}
\label{sec:limits}

OpenIIR cannot retrain its underlying LLM. The Multi-Generational type therefore evolves prompt templates and retrieval policies across generations, not model weights, and we are explicit about that scope. The Panel type carries the usual risks of LLM persona simulation: a participant whose simulated self is read at the workshop can disagree with what their persona said, and we treat that gap as a feature of the format rather than a defect, by running a participant validation pass after each panel. The Curated-Feed type uses a stylised belief vector, not a calibrated cognitive model; we report results as relative comparisons across rankers, not as absolute belief shifts. We also note that human relevance judgments themselves vary across annotators and time~\cite{parry2025variations}, which limits how tightly any synthetic-evaluation comparison should be interpreted.

Table~\ref{tab:roadmap} sketches modular extensions to the platform. Each row names a possible scenario type and the new primitive it would add beyond the core. Two rows (Feedback-Outage and Twin-Ecosystem) are in active development.

\section{Conclusion}
\label{sec:conclusion}

IR research now uses several distinct kinds of multi-agent simulation, and the field has been building one custom simulator per study. OpenIIR proposes a different default: a small shared core, a typed interface, and a small number of well-built scenario types that cover today's most common simulation styles. The Panel type is field-validated. The Social-Media, Curated-Feed, and Multi-Generational types are released with reference runs. Six modular extensions are sketched in Table~\ref{tab:roadmap}. The platform is open, local-first, and runs on a single laptop with no API key required. 

\bibliographystyle{ACM-Reference-Format}
\bibliography{sample-base}

\end{document}